\documentclass[pra,aps,twocolumn,nopacs,superscriptaddress,nofootinbib,reprint]{revtex4}
\usepackage{graphicx}  
\usepackage{dcolumn}  
\usepackage{bm}           
\usepackage{amsmath}
\usepackage{epsfig}
\usepackage{indentfirst}
\usepackage{psfrag}
\usepackage{subfigure}
\usepackage{amssymb}
\usepackage{tabularx,multirow}
\usepackage{color}
\usepackage[colorlinks,linkcolor=blue,citecolor=blue,urlcolor=blue,hyperindex,driverfallback=dvipdfm]{hyperref}
\usepackage[T1]{fontenc}
\usepackage{wasysym}

\def\rb{{\bf r}}      
      
\def\kb{{\bf k}}    
    
\def\Eb{{\bf E}}

\begin{document}
\title{Near-field dynamical Casimir effect}
\author{Renwen~Yu}
\affiliation{Department of Electrical Engineering, Ginzton Laboratory, Stanford University, Stanford, CA, USA}
\author{Shanhui~Fan}
\affiliation{Department of Electrical Engineering, Ginzton Laboratory, Stanford University, Stanford, CA, USA}
\begin{abstract}
We propose the dynamical Casimir effect in a time-modulated near-field system at finite temperatures. The system consists of two bodies made of polaritonic materials, that are brought in close proximity to each other, and the modulation frequency is approximately twice the relevant resonance frequencies of the system. We develop a rigorous fluctuational electrodynamics formalism to explore the produced Casimir flux, associated with the degenerate as well as non-degenerate two-polariton emission processes. We have identified flux contributions from both quantum and thermal fluctuations at finite temperatures, with a dominant quantum contribution even at room temperature under the presence of a strong near-field effect. We have found that the Casimir flux can be generated with a smaller modulation frequency through higher-order dynamical Casimir effect. We have conducted a nonclassicality test for the total radiative flux at finite temperatures, and shown that nonclassical states of emitted photons can be obtained for a high temperature up to $\sim 250\,$K.
Our findings open an avenue for the exploration of dynamical Casimir effect beyond cryogenic temperatures, and may be useful for creating tunable nanoscale nonclassical thermal states.
\end{abstract}
\maketitle


The exploration of fluctuation-induced phenomena in the near-field systems, where at least two bodies are brought in close proximity to each other, has attracted significant interests. Prominent effects in such near-field systems include quantum or thermal Casimir forces \cite{M94,L1956,DLP1961,MR98,BMM01,BCO02,KMM09,MCP09,PMG1986}, where mechanical forces between the bodies can be generated due to quantum or thermal fluctuations, and near-field radiative heat transfer \cite{PV1971,CG99,JMM05,KMP05,NSC08,BB14,OLF10,KHZ12,RSJ09,KSF15,ZGZ17,ZF16,SLL15,BEK17,BBJ11,AK22,paper286} with the heat transfer coefficient between the bodies significantly exceeding the far-field limit. Moreover, while most of the studies on near-field systems focus on systems that are static, in recent years there have been emerging interests in investigating fluctuation physics under the presence of time modulation \cite{KPJ15,KLR15,BLF20,SRJ21,VL23,ODP23,YF23,YF24}, where the material permittivity is varied as a function of time, with potential applications in cooling and energy harvesting \cite{YF24}. In all existing works on time-modulated near-field systems, however, the modulation frequency chosen is significantly smaller compared with the relevant resonance frequencies of the system. 

In this work, we consider a time-modulated two-body system composed of two different polaritonic materials, with a modulation frequency that is approximately twice the frequencies of relevant photonic modes supported in the system. We develop a rigorous fluctuational electrodynamics formalism to study both  quantum and thermal contributions in the near-field radiative flux between the two bodies. We show that in our time-modulated system the quantum component can dominate over thermal components even at room temperature provided that a strong near-field effect is available. Near-field enhanced degenerate as well as non-degenerate two-polariton emissions can be achieved by tuning the modulation frequency. We also show that the two-polariton emission can be produced with a smaller modulation frequency through higher-order processes. Furthermore, we consider a nonclassicality test for the total radiative flux at finite temperatures, showing that a nonclassical state of emitted photons can be obtained for a temperature up to $\sim 250\,$K. Our findings open an avenue for the exploration of dynamical Casimir effect beyond cryogenic temperatures, and may be useful for creating squeezed thermal vacuum for controlling atom-vacuum interactions \cite{DK97}, advancing quantum sensing \cite{PNL23}, and constructing nonclassical heat engines \cite{KFI17}. 

Our work is related to the dynamical Casimir effect that explores photon pair generations from quantum vacuum fluctuations using moving systems \cite{M1970_1,DKN93,DNM11}. The resulted photon flux is known as Casimir radiation or Casimir flux \cite{KG99,KN02}. The existing experimental demonstration of dynamical Casimir effect was carried out in cryogenic temperatures \cite{WJP11}.  Exploring dynamic Casimir effect at higher temperatures is of importance in the context of significant interests in exploring the interplay between quantum and thermal fluctuations in Casmir physics \cite{L1956,L97_2,YSM08,RWM10,OWA07}. The dynamical Casimir effect is also connected to vacuum frictions \cite{P97,paper157,paper166,paper199} and the Unruh effect \cite{U1976,NJB12}. Also, photon-pair generations have been considered in time-varying media \cite{L94_1,SRJ21,LCC07,CBC05,BH16}. Such pair generation can lead to the production of two-mode squeezed states \cite{JJW10,WJP11,JJW13} which is important for many quantum applications \cite{RL98,PCK92,AAA13}. However, the implication of such pair generation has not been previously studied in time-modulated near-field systems.

\begin{figure}
\noindent \begin{centering}
\includegraphics[width=0.5\textwidth]{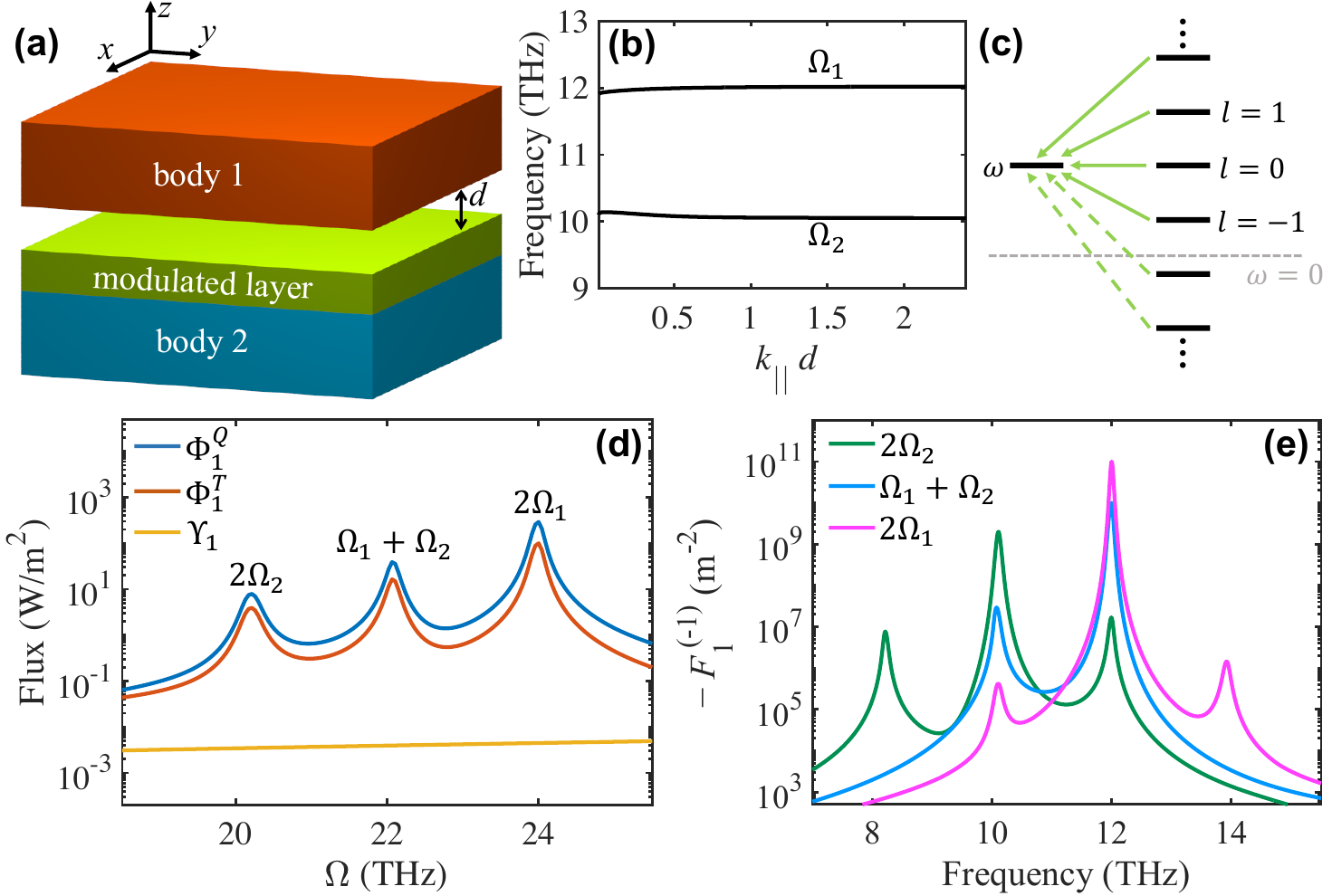}
\par\end{centering}
\caption{(a) Schematic of a planar photonic system composed of two semi-infinitely extended structures separated by a vacuum gap of a separation distance $d=10\,$nm. (b) Dispersion relations of the phonon polartions sustained in the system shown in panel (a). (c) Time modulation, of a modulation frequency $\Omega$, can enable frequency conversion processes from multiple source frequencies $\omega_l=\omega+l\Omega$ to the output frequency $\omega$, with $l$ an integer. (d) Net Casimir fluxes $\Phi_1^{Q}$ (quantum contribution) and $\Phi_1^{T}$ (thermal contribution), as well as inelastic scattering flux $\Upsilon_1$, received by body 1 as a function of the modulation frequency $\Omega$ at a temperature $T=300\,$K. (e) Time-modulated photon number flux spectra $-F_1^{(-1)}$ for three different modulation frequencies which are resonance frequencies labeled in panel (d). The frequency here is for the received photons at body 1.}
\label{Fig1}
\end{figure}

We consider a time-modulated photonic system composed of two planar structures separated by a vacuum gap of a distance $d=10\,$nm. The upper structure consists of a semi-infinite quartz (body 1, red region) substrate, whereas in the bottom structure a lossless time-modulated layer (green region) with a thickness of 22\,nm is located on top of a semi-infinite indium phosphide (InP, body 2, blue region) substrate, as shown in Fig.\ \ref{Fig1}(a). Both quartz and InP support phonon polaritons, and their permittivities are given by \cite{DMD12,P1985} $\epsilon_{1,2}(\omega)=\epsilon_{1,2}^{\infty}\left(1+\frac{(\omega_{1,2}^{L})^2-(\omega_{1,2}^{T})^2}{(\omega_{1,2}^{T})^2-\omega^2-i\gamma_{1,2}\omega}\right)$ with $\hbar \omega_{1}^{L}=$50\,meV, $\hbar \omega_{2}^{L}=$43\,meV, $\hbar \omega_{1}^{T}=$49\,meV, $\hbar \omega_{2}^{T}=$38\,meV, $\hbar\gamma_1=$0.26\,meV, $\hbar\gamma_2=$0.43\,meV, $\epsilon_1^{\infty}=$2.4, and $\epsilon_2^{\infty}=9.6$. The permittivity of the time-modulated layer is given as $\epsilon_3(t)=\epsilon_s+\delta\epsilon\,{\rm cos}(\Omega t)$, with $\epsilon_s=4$ the static permittivity, $\Omega$ the modulation frequency, $\delta\epsilon$ the modulation strength, and $t$ the time. The whole system is translationally invariant in the $x-y$ plane and is assumed to be in thermal equilibrium at a temperature $T$. 

Both bodies 1 and 2 support surface phonon polariton modes in the infrared frequency range. The dispersion relation of these modes features two rather flat bands, extending over a broad range of in-plane wavevectors $\kb_\parallel \equiv (k_x, k_y)$, with frequencies near $\Omega_1$ and $\Omega_2$, as shown in Fig.\ \ref{Fig1}(b). The modes at $\Omega_1$ and $\Omega_2$ are mostly confined in body 1 and 2, respectively. In our system, unless otherwise noted, we choose the modulation frequency $\Omega$ to be near the range of $[2\Omega_1,\,2\Omega_2]$. In general, dynamical Casimir effect manifests in terms of the emission of pairs of photons due to modulation. Here, the emission should be strongly enhanced by the presence of the surface phonon polariton \cite{RCN19,SRJ21}.

To illustrate the dynamical Casimir effect in our system, we focus on the net radiative energy flux $Q_1$ received by body 1. In the absence of modulation, we have $Q_1 = 0$ since the system is in thermal equilibrium. Following a fluctuational electrodynamics formalism \cite{YF23,YF24} for time-modulated radiative energy transfer, $Q_1$ can be decomposed into three components as $Q_1=\Phi_1^{Q}+\Phi_1^{T}+\Upsilon_1$ in the presence of modulation,
where  $\Phi_1^{Q}$ is the quantum contribution in the Casimir flux,  $\Phi_1^{T}$ is the thermal contribution in the Casimir flux, and $\Upsilon_1$ is the inelastic scattering flux, which has only thermal contribution. These components are given as
\begin{align}
	\Phi_1^{Q}&= -\int_0^{+\infty}  d\omega \sum_{l} u(-\omega_l)\hbar \omega  F_{1}^{(l)}(\omega), \label{eq:c1q} \\
	\Phi_1^{T}&=- \int_0^{+\infty}  d\omega \sum_{l}  u(-\omega_l)\hbar \omega  \left[n(\omega)+n(-\omega_l)\right]F_{1}^{(l)}(\omega), \label{eq:c1t} \\
	\Upsilon_1&=\int_{0}^{+\infty} d\omega \sum_{l} u(\omega_l)\hbar\omega  \left[n(\omega_l)-n(\omega)\right] F_{1}^{(l)}(\omega), \label{eq:up1}
\end{align}
where $u(\omega)$ is the Heaviside step function, $n(\omega)=\left[{\rm exp}(\hbar \omega/k_{\rm B}T)-1 \right]^{-1}$ is the Bose-Einstein distribution function, $\omega_l=\omega+l\Omega$ (with $l$ an integer) is the source frequency, and
\begin{align}
	&F_{\beta}^{(l)}(\omega)=\sum_{\alpha,i,j}\frac{2\epsilon_0^2}{\pi A}\epsilon''_{\beta}(\omega)   \epsilon''_{\alpha}(\omega_l) \nonumber \\
	&\times \int d\rb_{\beta} \int d\rb_{\alpha} \left| G_{ij}(\rb_{\beta},\rb_{\alpha};\omega,\omega_l) \right|^2 \label{eq:F12}
\end{align}
with $\alpha,\beta=1,2$ for body 1 or 2, $\epsilon''_{\alpha,\beta}(\omega)={\rm Im}\left \{ \epsilon_{\alpha,\beta}(\omega) \right\} $, $\rb_{\alpha,\beta}$ the spatial coordinates occupied by body $\alpha$ or $\beta$, and $A$ the surface area of the entire structure in the $x-y$ plane. In Eq.\ \ref{eq:F12}, $G_{ij}(\rb_{\beta},\rb_{\alpha};\omega,\omega_l)$, with $i,j=x,y,z$, is the Green's function for our time-modulated system. The quantity $F_{\beta}^{(l)}$, given in Eq.\ \ref{eq:F12}, is the photon number flux spectrum of the frequency conversion process from $\omega_l$ to $\omega$. In particular, when $\omega_l$ is negative, time modulation introduces mixing between positive (i.e., $\omega$) and negative (i.e., $\omega_l$) frequencies, which results in photon pair generations \cite{DKM90,DK92}. These conversion processes are indicated by dashed arrows in Fig.\ \ref{Fig1}(c). Eqs.\ \ref{eq:c1q} and \ref{eq:c1t} describe such pair generation processes, where the photons are generated from quantum and thermal fluctuations, respectively. Other processes, corresponding to the inelastic scattering of existing photons, are included in the inelastic scattering flux [indicated by solid arrows in Fig.\ \ref{Fig1}(c)], given by Eq.\ \ref{eq:up1}, which sources only from thermal fluctuations.

\begin{figure}
\noindent \begin{centering}
\includegraphics[width=0.5\textwidth]{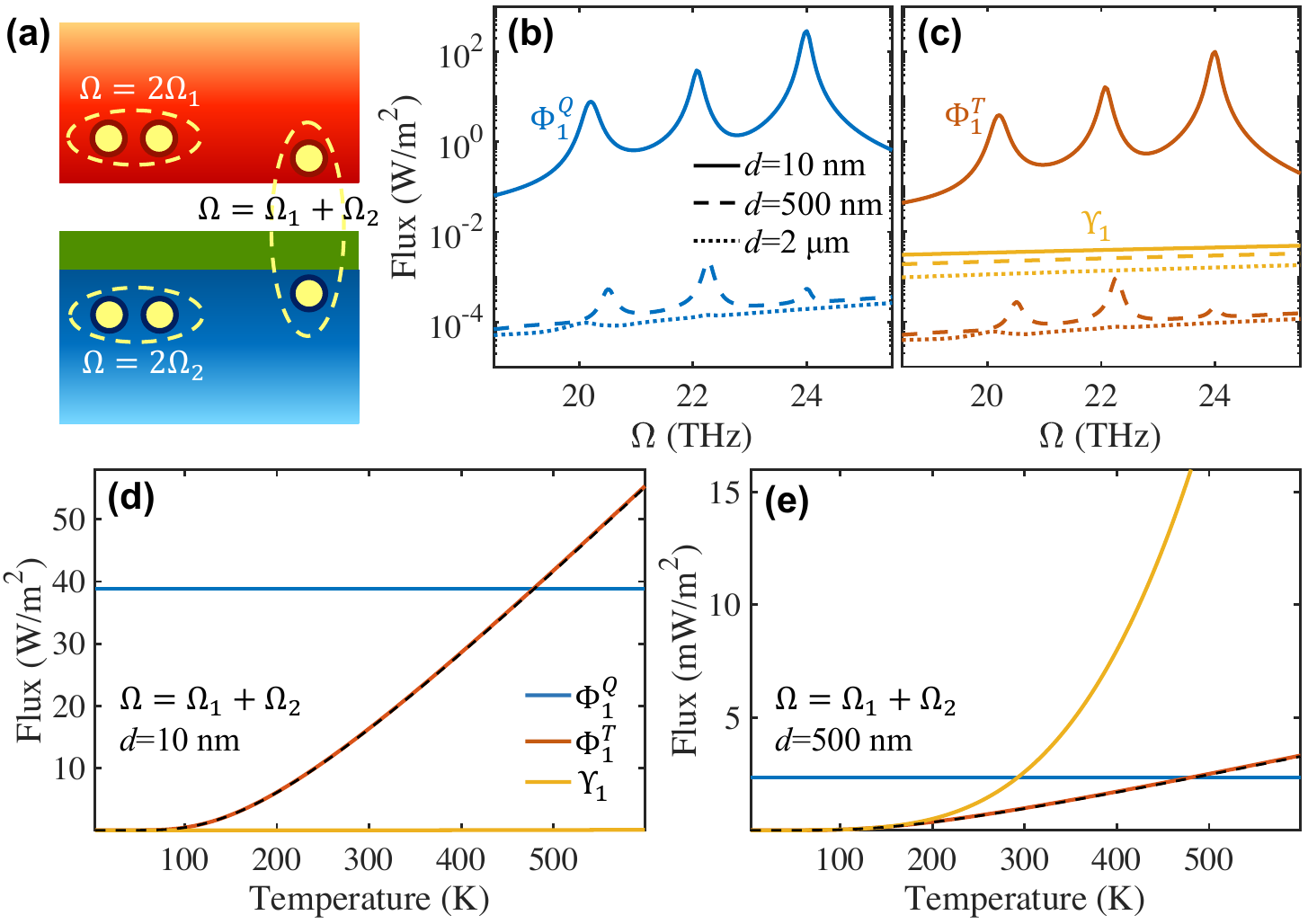}
\par\end{centering}
\caption{(a) Depending on the resonance condition, a pair of polaritons can be emitted with both of them into either body 1 or 2, or with one into body 1 and the other into body 2. (b,c) Flux components $\Phi_1^{Q}$, $\Phi_1^{T}$, and $\Upsilon_1$ as a function of the modulation frequency $\Omega$ with different separation distances $d$ at $T=300\,$K, where the quantum component ($\Phi_1^{Q}$) is shown in panel (b) and the thermal components ($\Phi_1^{T}$ and $\Upsilon_1$) are shown in panel (c). (d,e) Flux components $\Phi_1^{Q}$, $\Phi_1^{T}$, and $\Upsilon_1$ as a function of temperature $T$ with $d=10\,$nm (d) and $d=500\,$nm (e). The modulation frequency is assumed to be $\Omega_1+\Omega_2$. Black dashed curves are given by $\Phi_1^{Q}\times\left[n(\Omega_1)+n(\Omega_2) \right]$.}
\label{Fig2}
\end{figure}

In Fig.\ \ref{Fig1}(d), we show $\Phi_1^{Q}$, $\Phi_1^{T}$, and $\Upsilon_1$ as a function of the modulation frequency $\Omega$ at $T=300\,$K. The modulation strength is assumed to be $\delta\epsilon=0.4$ throughout this work unless otherwise specified. Three peaks, located at frequencies $\Omega\approx 2\Omega_2$, $\Omega_1+\Omega_2$, and $2\Omega_1$, can be found in $\Phi_1^{Q}$ (blue curve), each corresponding to a resonance condition for two-polariton generations. Specifically, at $\Omega\approx2\Omega_2$ ($\Omega\approx2\Omega_1$), degenerate generations of two polaritons of the same frequency $\Omega_2$ ($\Omega_1$) are resonantly enhanced. In contrast, at $\Omega\approx\Omega_1+\Omega_2$, non-degenerate generations of two polaritons of two different frequencies $\Omega_1$ and $\Omega_2$ are resonantly enhanced. Similar resonance features can be also seen in the thermal contribution $\Phi_1^{T}$ (red curve) of the Casimir flux, since these two terms share the same pair generation mechanism. In contrast, $\Upsilon_1$ (orange curve) is featureless, and its magnitude is much smaller than that of the Casimir flux, since the modulation frequency chosen are far from the resonance condition required for the inelastic scattering. 

We show in Figs.\ \ref{Fig1}(e) the spectra of $-F_1^{(-1)}$ (with $l=-1$), representing the most dominant process. For $\Omega=2\Omega_2$, we find three peaks with the dominant one in the middle around the frequency $\Omega_2$ (green curve), indicating the dominance of the resonant degenerate two-polariton emission at $\Omega_2$. The frequencies of the other two minor peaks are located at $\Omega_1$ and $2 \Omega_2 - \Omega_1$, respectively, corresponding to a non-degenerate two-polariton emission process. This process is weaker since there is no polariton mode near the frequency of $2\Omega_2 - \Omega_1$. The spectrum for $\Omega = 2\Omega_1$ (magenta curve) can be similarly interpreted. When $\Omega=\Omega_1+\Omega_2$, only two peaks, located around $\Omega_1$ and $\Omega_2$, can be seen, corresponding to a resonant non-degenerate two-polariton emission into the two polariton modes at $\Omega_1$ and $\Omega_2$, respectively.

In our system, when $\Omega = 2 \Omega_1$ or $2 \Omega_2$, we expect emission of polariton pairs mainly in body 1 or 2, respectively. In contrast, when $\Omega = \Omega_1 + \Omega_2$, we expect one polariton in the pair should be in body 1 and the other in body 2, as illustrated in Fig.\ \ref{Fig2}(a). Moreover, since these polaritons are strongly confined in either body, we expect the generation process should depend strongly on the separation distance between the bodies. In Fig.\ \ref{Fig2}(b), we show $\Phi_1^Q$ as a function of the modulation frequency $\Omega$ with different separation distances $d$. The results for $d=10\,$nm (solid curve) are taken from Fig.\ \ref{Fig1}(d). When increasing the separation distance to $d=500\,$nm, one can still observe three peaks but with drastically reduced magnitudes (dashed curve), corresponding to a much weaker resonant enhancement in the two-polariton emission. When further increasing $d$ to $2\,\mu$m, the peak features largely disappear, thus demonstrating a strong near-field effect in producing two-polariton emission into body 1. Note that a frequency shift of the resonance peaks can be seen when varying $d$ because the polariton dispersion changes as $d$ varies. The dependence of the flux magnitude on the separation distance shows the important role of the extremely subwavelength polariton modes in the near-field regime to provide large local density of states. As shown in Fig.\ \ref{Fig2}(c), similar observations can be found in $\Phi_1^T$ (red curves) as $d$ changes at a temperature of $T=300\,$K. Here, at room temperature, the magnitude of $\Phi_1^T$ is smaller than that of $\Phi_1^Q$. In contrast, $\Upsilon_1$ decreases more mildly as $d$ increases and shows no resonance features for all separation distances, as shown by the orange curves in Fig.\ \ref{Fig2}(c). As mentioned above, here the inelastic scattering is an off-resonance process. Thus the polariton response, which dominates in the near field, is not the key contributing factor here. Noticeably, $\Upsilon_1$ dominates over both $\Phi_1^Q$ and $\Phi_1^T$ at a large separation distance $d=2\,\mu$m.

In Figs.\ \ref{Fig2}(d,e), we show $\Phi_1^{Q}$, $\Phi_1^{T}$, and $\Upsilon_1$ as a function of temperature $T$ with $\Omega=\Omega_1+\Omega_2$ for two different separation distances. For $d=10\,$nm [Fig.\ \ref{Fig2}(d)], in the temperature range under study, $\Upsilon_1$ (orange curve) is much smaller than the other two components, as also demonstrated in Fig.\ \ref{Fig1}(d). $\Phi_1^Q$ (blue curve) is independent on the temperature because it originates from quantum fluctuations, whereas $\Phi_1^T$ (red curve) shows strong temperature dependence and becomes dominant over $\Phi_1^Q$ only when $T \apprge 480\,$K. These results demonstrate that the quantum contribution $\Phi_1^Q$ can dominate over all other contributions in the near-field regime. In contrast, when increasing the separation distance to $d=500\,$nm, $\Upsilon_1$ can be dominant over both $\Phi_1^Q$ and $\Phi_1^T$ when $T\apprge 292\,$K, as shown in Fig.\ \ref{Fig2}(e), and $\Phi_1^Q$ is no longer dominant. Noticeably, as shown in Figs.\ \ref{Fig2}(d,e), $\Phi_1^T$ can be simply reproduced by $\Phi_1^{Q}\times\left[n(\Omega_1)+n(\Omega_2) \right]$, given by the black dashed curves, which highlights the underlining mechanism of non-degenerate two-polariton emissions for both $\Phi_1^Q$ and $\Phi_1^T$. Also, as shown in Figs.\ \ref{Fig2}(d,e), both $\Phi_1^{T}$ and $\Upsilon_1$ vanish at zero temperature because they both source from thermal fluctuations.

\begin{figure}
\noindent \begin{centering}
\includegraphics[width=0.5\textwidth]{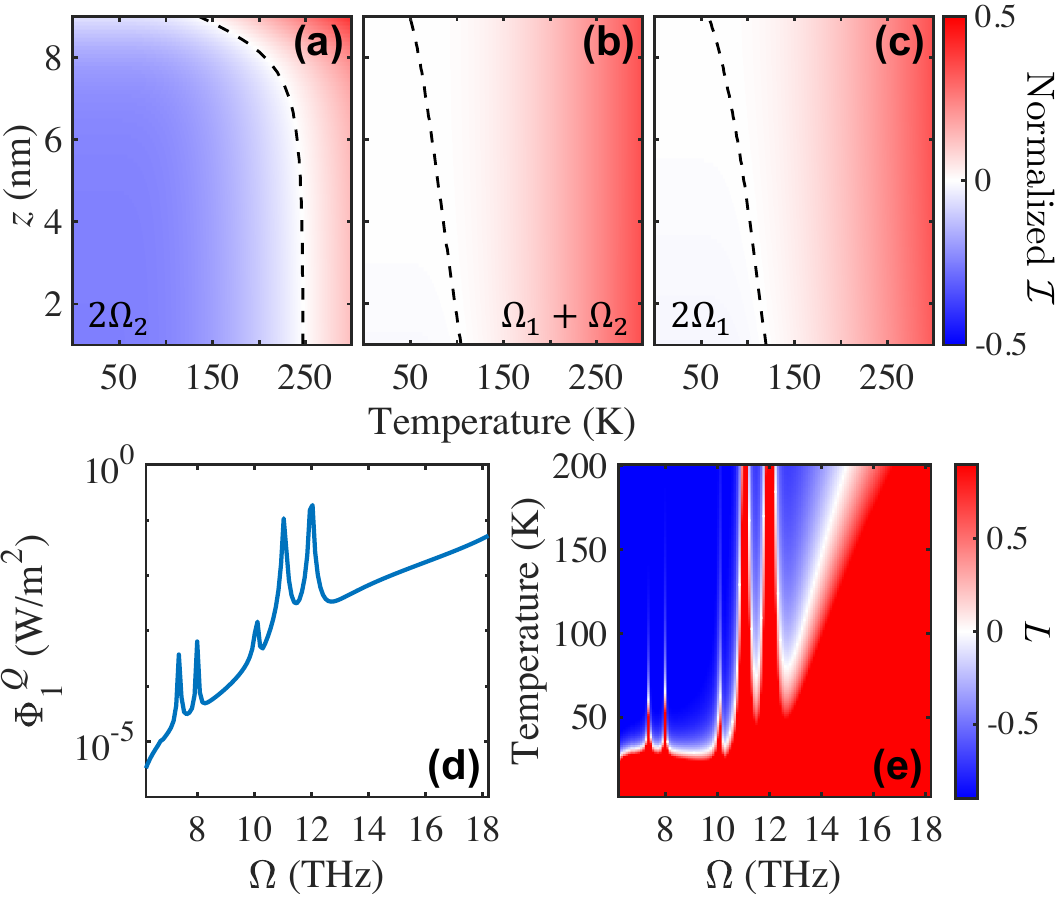}
\par\end{centering}
\caption{Normalized quantum-classical indicator $\mathcal I$ as a function of temperature $T$ and vertical distance $z$ measured from the top of the time-modulated layer, for the system considered in Fig.\ \ref{Fig1}. Normalized $\mathcal I$ is evaluated within the vacuum gap for three different modulation frequencies $\Omega=2\Omega_2$ (a), $\Omega_1+\Omega_2$ (b), and $2\Omega_1$ (c), as indicated in Fig.\ \ref{Fig1}(d). Black dashed curves indicate $\mathcal{I}=0$. The regions of nonclassical radiations locate on the left-hand side of the black dashed curves, where $\mathcal{I}<0$. (d) $\Phi_1^{Q}$ at lower modulation frequencies $\Omega$, as an extension of the results shown in Fig.\ \ref{Fig1}(d). (e) The relative strength of the quantum versus thermal contributions (defined as $L=|\Phi_1^Q |/|\Phi_1^T+\Upsilon_1 |-1$) as a function of temperature and modulation frequency.}
\label{Fig3}
\end{figure}

In the dynamical Casimir effect, the generation of correlated photon pairs can lead to two-mode squeezing \cite{WJP11}. Two-mode squeezed states can also be generated with thermal vacuum \cite{L90}. Inspired by these works, here we study the quantum nature of the emitted radiative flux by conducting a nonclassicality test for our system shown in Fig.\ \ref{Fig1}. For a Hermitian operator $\hat {\bf h}^{\dagger}\cdot\hat {\bf h}$, with $\hat {\bf h}$ composed of creation and annihilation operators, one can show, by using the Glauber-Sudarshan \textit{P} function, that any classical state of fields satisfies \cite{MBW10} $\langle: \hat {\bf h}^{\dagger}\cdot\hat {\bf h}:\rangle \geq 0$, where $::$ denotes normal ordering. Here for $\hat {\bf h}$ we consider the two-mode quadrature operator $\hat {\bf h}_\theta$, which is given as \cite{MBW10}
\begin{align}
	\hat {\bf h}_\theta=&\frac{1}{\sqrt{8\pi\epsilon_0\hbar}}\left[e^{i\theta}\hat{\Eb}(\rb,\omega_{-})+e^{-i\theta}\hat{\Eb}^{\dagger}(\rb,\omega_{-}) \right. \nonumber \\
	&\left.+ie^{i\theta}\hat{\Eb}(\rb,\omega_{+})-ie^{-i\theta}\hat{\Eb}^{\dagger}(\rb,\omega_{+}) \right]. \label{eq:ht}
\end{align}
In Eq.\ \ref{eq:ht}, $\theta$ is the angle that defines the principal squeezing axis, $\hat{\Eb}(\rb,\omega_{-})$ and $\hat{\Eb}(\rb,\omega_{+})$ represent the electric field operators for the two modes evaluated at two frequencies $\omega_{-}=\frac{\Omega}{2}-\delta\omega$ and $\omega_{+}=\frac{\Omega}{2}+\delta\omega$, respectively, with $\omega_{-}+\omega_{+}=\Omega$. The variable $\rb$ represents the spatial coordinate inside the vacuum gap. Note that for $\delta\omega \rightarrow 0$, $\hat {\bf h}_\theta$ can be used to characterize the degenerate two-mode squeezing (i.e., $\omega_{-}=\omega_{+}$). We define the quantum-classical indicator $\mathcal I$ as \cite{MBW10}
\begin{align}
	\mathcal{I}=\min_{\theta}\langle: \hat {\bf h}^{\dagger}_{\theta}\cdot\hat {\bf h}_{\theta}:\rangle 
	=C(\rb,T)-|B(\rb,T)|.   \label{eq:hh}
\end{align}    
Here $C(\rb,T)=\sum_{M=\pm} R(\rb;\omega_M,\omega_M)n(\omega_M)$ and $B(\rb,T)=\sum_{M=\pm} R(\rb;\omega_{M}-\Omega,\omega_M)[2n(\omega_M)+1]$
with $R(\rb;\omega,\omega')=\sum_{\alpha,i,j}\epsilon''_{\alpha}(\omega') \times\int d\rb_{\alpha}{G}_{ij}(\rb,\rb_{\alpha};\omega,\omega') {G}^{*}_{ij}(\rb,\rb_{\alpha};\omega',\omega')$,  
where the superscript * denotes the complex conjugate operation. Note that $C(\rb,T)$ is a real positive quantity.

In Eq.\ \ref{eq:hh}, $C(\rb,T)$ arises from the inelastic scattering flux, whereas $B(\rb,T)$ arises from the Casimir flux, containing both thermal and quantum components, the presence of which is essential to reach the nonclassicality condition $C(\rb,T)<|B(\rb,T)|$,
leading to $\mathcal{I}<0$. In Figs.\ \ref{Fig3}(a-c), we show  the normalized quantum-classical indicator $\mathcal{I}/{\mathcal {N}}$, with ${\mathcal N}=\sum_{M=\pm} R(\rb;\omega_M,\omega_M)/2$, as a function of temperature $T$ and vertical distance $z$ measured from the top of the time-modulated layer, for the system considered in Fig.\ \ref{Fig1}. The normalized quantum-classical indicator is evaluated within the vacuum gap for three different modulation frequencies (see labels inside). In Figs.\ \ref{Fig3}(a) and (c), we have $\delta\omega=0$ for degenerate two-mode squeezing, whereas $\delta\omega=(\Omega_1-\Omega_2)/2$ for non-degenerate two-mode squeezing in Fig.\ \ref{Fig3}(b). As shown in Figs.\ \ref{Fig3}(a-c), the state of the emitted photon flux can exhibit nonclassical behaviors even at finite temperatures. The behaviors change from nonclassical to classical when increasing the temperature for all three modulation frequencies due to an increased thermal noise. Moreover, there is also a spatial dependence in the state of the radiation. Away from the time-modulated layer (i.e, as $z$ increases), the nonclassical behavior is weaker since the effect of modulation is weaker, as shown in Figs.\ \ref{Fig3}(a-c). Moreover, when $\Omega=2\Omega_2$ [Fig.\ \ref{Fig3}(a)], there is a larger parameter range where nonclassical behaviors can be observed, since the time-modulated layer is attached to body 2 and the degenerate two-mode squeezing at $\Omega_2$ is more efficient. In this case, the nonclassical behavior can survive up to a temperature of $\sim250\,$K. 

Interestingly, due to the near-field effect, higher-order dynamical Casimir effects can occur in our system, where two-polariton generation can occur at lower modulation frequencies, as shown in Fig.\ \ref{Fig3}(d). The three resonance peaks around $2\Omega\sim2\Omega_{1,2}$ correspond to the second-order process, whereas the resonance peaks around $3\Omega\sim2\Omega_{1,2}$ correspond to the third-order process. Furthermore, in Fig.\ \ref{Fig3}(e), we show the relative strength of the quantum versus thermal contributions (defined as $L=|\Phi_1^Q |/|\Phi_1^T+\Upsilon_1 |-1$) as a function of temperature and modulation frequency. When $L>0$, the flux is quantum dominant. Notably, the flux can be still quantum dominant at high temperatures with simultaneously a smaller modulation frequency, due to the presence of higher-order dynamical Casimir effect. Note that a dominant quantum contribution over the thermal ones at room temperature can be still obtained with a much smaller modulation strength ($\delta\epsilon=0.004$). Similar modulation conditions have been achieved experimentally by using an optically pumped Kerr medium as the time-modulated layer \cite{GHV24}.


In summary, we have shown that a strong dynamical Casimir effect, including higher-order processes, can be obtained in a time-modulated near-field polaritonic system composed of two bodies, which results in near-field enhanced degenerate and nondegenerate two-polariton generations. The two-polariton generation is manifested as a photon flux, named Casimir flux, the quantum component of which can dominate over all the other thermal components even at room temperature. We have shown that nonclasscical states of emitted photons can be obtained even for a high temperature up to $\sim 250\,$K.
Our findings may be of importance for developing tunable nanoscale nonclassical light sources, and may motivate further explorations for quantum applications based on thermal light.

This work has been supported by a MURI program from the U. S. Army Research Office (Grant No. W911NF-19-1-0279), and by the Department of Energy (Grant No. DE-FG02-07ER46426).


\end{document}